%% file: main.tex
\documentclass[sigconf]{acmart}
\usepackage{booktabs} 
\usepackage{mathtools} 
\usepackage{amsmath}

\newcommand{\dblp}{\ensuremath{\textsf{dblp}}\xspace}

\newcommand{\now}{\ensuremath{\textsf{NOW}}\xspace}
\newcommand{\magraph}{\ensuremath{\textsf{MAG}}\xspace}
\newcommand{\ndcg}{\ensuremath{\textsf{ndcg}}\xspace}

\usepackage{listings}
\usepackage{xcolor-material}
\usepackage{listing-set}

\usepackage[colorinlistoftodos,prependcaption,textsize=tiny]{todonotes}
\presetkeys%
    {todonotes}%
    {inline,backgroundcolor=yellow}{}

\usepackage{blindtext}
\usepackage{csvsimple}
\usepackage{diagbox}

 \setcopyright{none}

\copyrightyear{2018}
\acmYear{2018}
\acmConference[JCDL '18]{The 18th ACM/IEEE Joint Conference on Digital Libraries}{June 3--7, 2018}{Fort Worth, TX, USA}
\acmBooktitle{JCDL '18: The 18th ACM/IEEE Joint Conference on DigitalLibraries, June 3--7, 2018, Fort Worth, TX, USA}
\acmPrice{15.00}
\acmSubmissionID{jcdl090s}
\acmDOI{10.1145/3197026.3197069}
\acmISBN{978-1-4503-5178-2/18/06}

\begin{document}

\title{Prioritizing and Scheduling Conferences for Metadata Harvesting in dblp}

\input{input/authors}
\renewcommand{\authors}{Mandy Neumann, Christopher Michels, Philipp Schaer, and Ralf Schenkel}
\renewcommand{\shortauthors}{Mandy Neumann, Christopher Michels, Philipp Schaer, and Ralf Schenkel}


\begin{abstract}
Maintaining literature databases and online bibliographies is a core responsibility of metadata aggregators such as digital libraries.
In the process of monitoring all the available data sources the question arises which data source should be prioritized.
Based on a broad definition of information quality we are looking for different ways to find the best fitting and most promising conference candidates to harvest next.
We evaluate different conference ranking features by using a pseudo-relevance assessment and a component-based evaluation of our approach.
\end{abstract}


%
%

%




\maketitle

\input{input/body}

\bibliographystyle{ACM-Reference-Format}
\bibliography{main}

\end{document}

%% file: input/authors.tex

\author{Mandy Neumann}
\orcid{0000-0003-3694-4997}
\author{Philipp Schaer}
\orcid{0000-0002-8817-4632}
\affiliation{%
  \institution{TH K{\"o}ln}}
\email{firstname.lastname@th-koeln.de}

\author{Christopher Michels}
\orcid{0000-0003-0523-8547}
\author{Ralf Schenkel}
\orcid{0000-0001-5379-5191}
\affiliation{%
  \institution{Trier University, \dblp}}
\email{(michelsc,schenkel)@uni-trier.de}

%% file: input/body.tex
\input{input/introduction}
\input{input/method}
\input{input/dataset}
\input{input/evaluation}
\input{input/discussion}
\input{input/acknowledgements}

%% file: input/introduction.tex
\section{Introduction}
%

Monitoring and maintaining databases and data sources are core responsibilities of metadata aggregators such as digital libraries.
Metadata provided by publishers at specific intervals might be overdue which needs to result in some notification.
Similarly, self-sufficient metadata harvesting relies on scheduling or event tracking and prognosis, for example.
Integrating such heterogeneous sources is naturally error-prone.
The various data sources involved need to be continuously monitored and maintained.

The database publisher \emph{dblp computer science bibliography} (\dblp) provides approx. 4 million publication records originating from more than 4,000 conferences and 1,500 journals.
These records are provided by sources varying from emminent commercial publishers to isolated alternating web sources (as of 01/2018).
Acquiring metadata from this extent of sources mainly involves data provided by publishers as well as highly customized harvesting software, so-called wrappers.
However, data sets might be provided at arbitrary intervals or wrappers might not be successful in retrieving the expected metadata. Several questions arise from these problems:
1)~Which data are expected next and when?
2)~Are they missing and overdue?
3)~Which source usually provides the metadata?
4)~Is there an alternative source?
5)~Is there a ready solution for harvesting the data?
6)~Which are the most urgent?
These questions illustrate the expected costs of finding a solution to missing data in the acquisition process.

Simply crawling every possible data source every night is no solution as the conceptional and technical costs are too high.
At the scale of \dblp, filtering data deliveries, leveraging metadata harvesting, and other challenges of database maintenance require some form of automated prioritization in order to be more efficient: Which missing data from which data source are the most urgent?

A similar issue is known for web crawlers where it is crucial to prioritize and rank the most promising web pages for the following web crawling process.
Different strategies have been proposed, such as pre-calculating a PageRank and sorting the crawling candidates according to \cite{DBLP:conf/www/Baeza-YatesCMR05}.
Other approaches include temporal web link-based rankings to estimate a better time-dependent web page authority compared to PageRank \cite{Dai:2010:Freshness}.
While these strategies help to achieve the objective of crawling important pages early in the process, applying these strategies for the use case of \dblp does not work as we miss the necessary information, such as the linking between conferences.
Recent work on \dblp data has implemented conference recommenders which cluster and rank conferences based on author-conference similarities and co-authorship networks \cite{DBLP:journals/jbcs/GarciaNLCL17}.

The research question is to find a prioritization mechanism for conference series in order to get a ranked list with regard to their expected urgency for the data acquisition process.
A highly ranked conference in this list is a conference which the \dblp editorial staff should include in the next data harvesting or import routine.
The need for smart, prioritized metadata harvesting is most prominent with conference series and their dynamic event structure in comparison with journals or other publication venues.

Based on a broad definition and idea of information quality and the question which features are necessary to measure the goodness or quality of a data source \cite{DBLP:journals/tcdl/Ayala17, DBLP:conf/sigir/ZhuG00} we are looking for different ways to find the best fitting and most promising conference candidates.
We evaluate different conference ranking features by using a pseudo-relevance assessment and a component-based evaluation of our approach.

%% file: input/method.tex
\section{Ranking of Conference Candidates}
\label{sec:ranking}
Prioritizing which conference is checked next for new proceedings worth adding to a literature database at a given date is a typical ranking problem: the set of conferences needs to be ranked in descending order according to their relevance for the database.

The following notations pertain to the description and evaluation of all scoring factors for the ranking of conferences:
We consider the set of conferences $C$, where a conference (JCDL, SIGIR, etc.) $c\in C$ is a set of conference events (events) $E(c)=\{e_1,\ldots,e_n\}$ archived in the bibliography.
The date of an event is denoted by $d(e)$.
Dates $d=(m,y)$ are represented as pairs of a month $m$ (numeric) and a year $y$.
The difference of two dates is the number of months between the dates.
Adding $\delta$ months to a date yields a date that is $\delta$ months later.
The function $year(d)$ yields the year, $month(d)$ yields the month of date $d$.
The current date is denoted by $\now$.

For each conference, the characteristic interval $\delta_{year}(c)$ is estimated as the median of distance of the years of the five most recent subsequent events in $E(c)$ (or of all events if $c$ has less than five events).
Additionally, we define the month $m(c)$ in which events usually take place as the most frequent (mode) $month(d(e))$ of all events in $E(c)$.
Finally, we determine the characteristic delay $\delta_{month}(c)$ of a conference as the median of the delays in months between the conference event and the entry of its proceedings into the database.
By analogy with $\delta_{year}(c)$, we use up to the five most recent events in $E(c)$ for calculation.
\subsection{Scoring Conference Candidates}\label{subsec:scoring}
In order to score and rank the inclusion candidates for conference events,
we define the following intuitive and comprehensible criteria.
Determining the relevance of a conference to a bibliography depends to a large extent on the fact that conference events usually occur on a regular basis.
Consequently, observing and evaluating historical data of database maintenance is crucial to a relevance-based ranking for conferences.
Additionally,
external regional conference ratings,
the internationality of a conference,
a citation score, and
a prominence score
are integrated as boosting factors in the relevance ranking for conferences,
whereas presumably discontinued conferences receive a penalizing factor.
We describe in section \ref{sec:dataset} how these scores can be computed for \dblp. 
Note that not all information necessary for these factors might be available for a given bibliographic database.

\textbf{Delay.}
From the entry date $d_n(c)$ of the last known event of a conference, we can estimate the expected entry date of the next event in this series $d_{n+1}(c)\ =\ (m(c),\ year(d_n(c))\ +\ \delta_{year}(c))\ +\ \delta_{month}(c)$.
For any given point in time, we can then determine $\Delta(c) = \now - d_{n+1}$
to see if and how long the conference $c$ is already presumably overdue.
We derive the following base scoring factor $w_{delay}(c)$ from $\Delta(c)$,
ranking conferences with a smaller delay, for which the expected next entries are more recent, higher than those with a greater delay. %
{%
\setlength{\abovedisplayskip}{3pt}
\setlength{\belowdisplayskip}{3pt}
\begin{equation}
  w_{delay}(c) = \begin{dcases}
   4 & 0 \leq \Delta(c) \leq 3 \\
   3 & 4 \leq \Delta(c) \leq 7 \\
   2 & 8 \leq \Delta(c) \leq 11 \\
   1 & 12 \leq \Delta(c) \\
   0 & \Delta(c) < 0
  \end{dcases}
  \label{eq:cases}
\end{equation}}
To calculate the score for a conference, this base scoring factor is multiplied by one of the following weighting factors.

\textbf{Conference ratings.}
Existing conference ratings such as \emph{CORE Computing Research \& Education}\footnote{\url{http://portal.core.edu.au/conf-ranks}} (CORE) rate conferences to model their importance for the research communities involved. Such external ratings might not cover the entire set of conferences for a given database because they are usually subject to a specific regional, scientific, or other type of focus.
Regardless of their scope, combining these ratings serves the purpose of boosting the most important conferences for all rating-specific perspectives.
If possible, existing rating classes are integrated and mapped to numeric classes.
Conferences without any rating values are attributed to the residue class $0$.
Then, $r(c)$ is the average of all numeric rating values attributed to $c$, with the corresponding weighting factor:
{%
\setlength{\abovedisplayskip}{3pt}
\setlength{\belowdisplayskip}{3pt}
\begin{equation}
w_r(c):=1+\frac{r(c)}{max_Cr(c)}
\end{equation}
}

\textbf{Internationality.}
We define the internationality of a conference $i(c)$ as the number of distinct countries where the conference has taken place.
The idea of this factor is that it reflects how often the conference venue is changing locations across countries.
We argue that conferences which are more international in this sense should be prioritized against local ones.
By analogy with the rating factor $w_r(c)$, we put $i(c)$ in relation to the whole collection to obtain the internationality weighting factor:
{%
\setlength{\abovedisplayskip}{3pt}
\setlength{\belowdisplayskip}{3pt}
\begin{equation}
  w_i(c):=1+\frac{i(c)}{max_Ci(c)}
\end{equation}
}

\textbf{Discontinued conferences.}
The development of conferences and their respective events over time constitutes a dynamic process.
Conferences might have been merged, or co-located with others, or they might be discontinued entirely.
Database publishers might have dismissed particular conferences, i.e., due to a change of scope.
In either case, it is impractical to rank these conferences high because they are dozens of months overdue.
Consequently, suspended conferences are reflected by the penalizing weighting factor:
{%
\setlength{\abovedisplayskip}{3pt}
\setlength{\belowdisplayskip}{3pt}
\begin{equation}
  w_d(c) := \left(1+{\frac{\#years~since~last~entry}{\delta_{year}(c)}}\right)^{-2}
\end{equation}
}

\textbf{Citations.}
In the same way incoming links account for the popularity of a web page \cite{DBLP:conf/sigir/ZhuG00}, the number of papers by which a given paper is cited can be used to measure its influence on the research community.
With $cit(e,y)$ denoting the number of incoming citations in year $y$ to all papers from event $e$ as well as the number of papers $p(e)$ for the same event, we compute the average number of citations per paper across the events of a conference $E(c)$ with $Y=\{y|\exists e\in E(c):y=year(d(e))\land y< year(\now)\}$:
{%
\setlength{\abovedisplayskip}{3pt}
\setlength{\belowdisplayskip}{3pt}
\begin{equation}
  w_{cit}(c)=1+\frac{cit(c)}{max_Ccit(c)},~cit(c)={\sum\limits_{y\in Y}^{}}~{\sum\limits_{e\in E(c)}^{}\frac{cit(e,y)}{p(e)}}
\end{equation}
}

\textbf{Author Prominence.}
An authors' scientific prominence is commonly measured by the number of their publications.
Thus, events with many frequently published scientists are preferred over events which have published rather unknown authors.
For each event $e$ we may know its number of distinct authors $a(e)$ and $p(a, y)$ as the number of records per author $a$ until year $y$.
Equation \ref{eq:prom-e} shows how the prominence score is computed for an individual event.
We define the author prominence weighting factor as in \ref{eq:prom-c}.
Events with less than 10 publications are excluded due to too much noise, resulting in the subset $\overline{E}(c) = \{e_i| e_i \in E(c)\land p(e_i) >= 10\}$ of $E(c)$. 
{%
\setlength{\abovedisplayskip}{3pt}
\setlength{\belowdisplayskip}{3pt}
\begin{gather}
    prm(e)=\frac{1}{a(e)}{\sum\limits_{i=1}^{a(e)} p(a_i,year(\now)-1)}\label{eq:prom-e}\\
    w_{prm}(c)=1+\frac{prm(c)}{max_Cprm(c)},~prm(c) = \frac{1}{|\overline{E}(c)|}{\sum\limits_{e\in\overline{E}(c)}^{}prm(e)}\label{eq:prom-c}
\end{gather}
}
\begin{table}[t]
\caption{Example values for conference features used in the rankings, computed for December 2016.}\label{table/example}
  \begin{center}
\begin{tabular}{@{}lrrrrrrr@{}}\toprule
  $c$ & $\Delta(c)$ & $w_{delay}$ & $w_r$ & $w_i$ & $w_d$ & $w_{cit}$ & $w_{prm}$\\\midrule
  jcdl  & 3   & 4 & 1.88 & 1.192  & 0.250   & 1.029  & 1.312\\
  tpdl  & 0   & 4 & 1.63 & 1.577  & 0.250   & 1.024  & 1.352\\
  icadl & 0   & 4 & 1.75  & 1.385  & 0.250   & 1.009  & 1.347\\
  dl    & 146 & 1 & 1.00   & 1.039   & 0.004  & 1.091 & 1.445\\ \bottomrule
\end{tabular}
  \end{center}
\end{table}

%% file: input/dataset.tex
\section{Data Sets}\label{sec:dataset}
The evaluation of these scoring factors for the conferences of \dblp is based on the following data sets.
A different literature database might require additional, compatible sources of information in order for the scoring factors above to be applied.
We explain how external ratings and a citation graph have been consolidated with the \dblp database to perform and evaluate the conference candidate scoring described above.

\textbf{\dblp.}
For each modification of a record the updated metadata is persisted in \dblp with a timestamp.
Thus, the entry date for each proceedings record is obtained from its earliest modification date.
Our analyses are based on the \dblp as of 2018/01/04, containing  more than 4 million distinct records (excluding author homepages).
Of those, about 36,000 records are proceedings of approx. 4,300 different conferences.
As the conferences' coverage varies among the computed weighting factors, a conference which is not covered by a specific factor receives the neutral weight $1$ for that factor.

Additional information for the evaluation of the weighting factors can be acquired from \dblp records.
All forms of the function $p()$ and $a()$ above constitute simple queries based on person, record, and conference indentifiers as well as time-range restrictions.
The country of an event is parsed from the geographical information\footnote{For parsing geographical information, the Python library geotext (\url{https://pypi.python.org/pypi/geotext}) by Yaser Martinez Palenzuela was used.} in the titles of proceedings.
Similarly, the date $d(e)=(m,y)$ of an event is extracted from its title if possible.
As this date information is crucial to our scorings, we filter the available data sets to retain only those 4,129 conferences where this information is available for at least one event record.

\textbf{Conference ratings.} The Australian rating CORE in 2008 and 2017 as well as the results of the conference rating in \cite{DBLP:conf/jcdl/MartinsGLP09} from 2008 serve as the basis for the conference rating weighting factor.
Their combined rating classes are $\{A*, A, B, C, Other\}$.
For $r(c)$ the rating values across the 3 ranked lists are mapped to their equivalent from the numeric rating classes $\{4,3,2,1,0\}$.
Thus, $r(c)$ is higher than 0 for 824 conferences.

\textbf{Citations.}
To enhance the dblp data set with citation information, 166,192,182 publication records from the \emph{Microsoft Academic Graph}\footnote{\magraph is a part of the Open Academic Graph by AMiner (\url{https://aminer.org/open-academic-graph})} (\magraph) and their relationships are mapped to their equivalents in \dblp.
About 3.6 million records occurring in \magraph were successfully matched based on their DOIs and titles.
To compute $cit(e,y)$ for 4,210 conferences with matching records in \dblp and \magraph, the incoming citations are counted for each conference and each year of citation starting from 1970.
These numbers are based on the \magraph as of 2017/06/09.

%% file: input/evaluation.tex
\section{Evaluation}
In order to evaluate the performance of our rankings, relevance judgments for all conferences are needed.
As it is impossible to get human judgments for thousands of conferences, we resorted to pseudo-relevance.
Our graded pseudo-relevance is interval-based, inverted and takes into account only the delay between the entry month of the latest record and the current month which we refer to as $\delta_{month}(c)$.
Conferences where the newest event had been added more recently to the dataset (small delay) receive a higher relevance rating than those with a larger delay.
The pseudo-relevance scores $rel_p(c)$ are assigned in the same way as the delay-based scoring factor $w_{delay}(c)$, see eq. \ref{eq:cases}, with $\delta_{month}(c)$ instead of $\Delta(c)$.

After establishing the relevance judgments, we perform a leave-out evaluation.
First, all the latest entries of 2016 are ignored.
For each month of 2016 and each conference, the base scoring factor $w_{delay}(c)$ is weighted by one of the five features discussed above.
Then, a priority ranking to indicate which conferences should be added next is predicted by sorting them by score in descending order.
A sliding window over the months of 2016 is used to calculate the qrels and ranking scores for each month, averaging the results over the year afterwards.
Performance measurements are carried out by the standard trec\_eval evaluation procedure.
Since we have a graded relevance scale we compute the normalized Discounted Cumulative Gain (\ndcg) on different cut-off levels.
As our evaluation year $year(\now)$ is set to 2016, all weighting factors discussed in section \ref{subsec:scoring} are calculated taking into account all events and records up to and including $year(\now)-1$ (in this case 2015).
Table \ref{table/example} shows some examples of the available data for scoring.

\begin{table*}[t]
\caption{Overview on \ndcg-100 values for each month and the year's average. }\label{table/resultsMonths}
  \begin{center}
	\begin{filecontents*}{tables/ndcg.csv}
system;jan;feb;mar;apr;may;jun;jul;aug;sep;oct;nov;dec;avg
baseline;0.240;0.338;0.353;0.434;0.510;0.505;0.566;0.601;0.690;0.632;0.581;0.605;0.505
conf. rating;0.230;0.378;0.524;0.627;0.682;0.781;0.783;0.769;0.758;0.725;0.746;0.736;0.645
internationality;0.226;0.331;0.507;0.610;0.639;0.679;0.735;0.758;0.716;0.681;0.732;0.679;0.608
discontinued;0.291;0.411;0.615;0.727;0.686;0.673;0.702;0.676;0.779;0.743;0.707;0.711;0.643
citations;0.225;0.333;0.442;0.517;0.546;0.583;0.621;0.634;0.741;0.714;0.652;0.643;0.554
prominence;0.248;0.423;0.568;0.637;0.601;0.621;0.671;0.682;0.752;0.678;0.718;0.696;0.608
\end{filecontents*}
\csvautobooktabular[separator=semicolon]{tables/ndcg.csv}
  \end{center}
\end{table*}

\begin{table}[t]
\caption{Comparison of \ndcg values on different cut-offs. Statistical differences to the baseline tested with two-sided t-test ($*** = p<0.001$, $** = p<0.01$, $ * = p<0.05 $).}\label{table/resultsCutoffs}
  \begin{center}
    \begin{filecontents*}{tables/ndcgcutoffs.csv}
system;ndcg-10;ndcg-20;ndcg-100;ndcg-200
baseline;0.530;0.545;0.505;0.439
conf. rating;0.739**;0.716**;0.645***;0.597***
internationality;0.616;0.632;0.608***;0.575***
discontinued;0.713**;0.686***;0.643***;0.594***
citations;0.588;0.575;0.554***;0.548***
prominence;0.681**;0.662**;0.608***;0.577***
\end{filecontents*}
\csvautobooktabular[separator=semicolon]{tables/ndcgcutoffs.csv}
  \end{center}
\end{table}

\subsection{Results}
The results of the evaluation are shown in tables \ref{table/resultsMonths} and \ref{table/resultsCutoffs}.
Table \ref{table/resultsMonths} gives an overview on all computed \ndcg-100 values for each month of the evaluation year and the year's average.
A hundred documents is a typical amount of conferences the dblp editors can handle in one month.
The raw delay $\Delta(c)$ of a conference constitutes the baseline.
We can see that each ranking factor outperforms our baseline, conference rating and the discontinuity being the best-performing factors.
Over the months of the year, we observe some fluctuation, which might be attributed to certain time slots where the number of simultaneous conference events peak.
In the beginning of the year, the \ndcg values are remarkably low.
The values peak in September and decline towards the end of the year.
Nevertheless, the overall trend regarding the individual ranking factors is quite stable over the year.

Table \ref{table/resultsCutoffs} also shows that differences between the rankings and the baseline are always highly significant for discontinued and conference rating factors.
These two factors are also performing best in all observations.
While for low cut-off values the differences between the rankings are high, they are lower for high cut-offs.
This pattern is plausible as divergences between different rankings balance out in the long tail of the ranked list.

%% file: input/discussion.tex
\section{Discussion and Future Work}
The idea of utilizing quality metrics in information retrieval systems is not new. Most of these metrics target the goodness of web sources \cite{DBLP:conf/www/Baeza-YatesCMR05}. In this work we transfer this idea to the question which of the missing metadata in a literature database needs to be included next. 


From a synoptic, philosophical perspective \cite{DBLP:journals/tcdl/Ayala17}, the most prominent features of information quality are accuracy, completeness, currency, compactness, usability, consistency, and credibility.
These seven facets also permeate the perspectives of other scientific fields, such as information and computer science.
Similar to \cite{DBLP:conf/sigir/ZhuG00}, we propose several features of conference metadata to improve the quality and maintenance of literature databases.
Our features operationalize the following quality metrics: \textit{currency}, through the delay of expected data and our proposed discontinuity factor; \textit{popularity} via citation, internationality and prominence scores; and \textit{credibility} via the conference ratings.
We have proposed an evaluation framework to evaluate the effectiveness of these operationalized features for the problem of ranking data sets according to their expected urgency for a data acquisition process.
We were able to show that the baseline ranking, which only takes the raw delay into account, can be outperformed by a number of semantically plausible factors.
When it comes to create a priority ranking of conference data sets, currency expressed through penalizing a delay with a measure for discontinuity, and credibility of a conference seem to be the best-performing factors.


We had to make some compromises to perform our analyses.
For a number of conferences, not all desirable information was available.
For example, if no venue is mentioned in the title of proceedings, or no citation numbers are available from MAG, we will ignore this data for scoring.
Our definition of internationality also needs to be revised to account for conferences which do not change locations frequently but still are of international relevance.
For further investigations, we will also need to separate conferences from workshops co-located with them.
In the current setting they are bundled together, which may add some noise to our data.




Our evaluation framework allows to perform follow-up investigations, for example by employing additional ranking factors that relate to the remaining quality metrics.
We also want to combine factors to come up with the best-performing solution for a specific use case.
In future work, we consider additional evaluation metrics to enhance our ranking.


%% file: input/acknowledgements.tex
\begin{acks}
This work was supported by the DFG (Project ID: 217852844).
We thank Hendrik Adam for his contribution to the implementations of the different ranking factors and the evaluation framework.
\end{acks}